\documentclass[preprint]{elsarticle}
\usepackage[utf8]{inputenc}
\usepackage{amsmath}
\usepackage{amsfonts}
\usepackage{amssymb}
\usepackage{graphicx}
\usepackage{physics}

\begin{document}
\begin{frontmatter}

\title{Evolution of magnetic fields in cosmic string wakes. }
\author{Soumen Nayak, Sovan Sau and Soma Sanyal}
\date{}
\address{School of Physics, University of Hyderabad, Gachibowli, Hyderabad, India 500046}


\begin{abstract}

{We study the evolution of magnetic fields in cosmic string wakes in a plasma with a  low resistivity. The initial magnetic field in the wake is modelled on the magnetic fields that are generated by the motion of particles around cosmic strings. The plasma is characterized by a high beta value. We find multiple shock like structures developing in the wake of the string. We study the detailed structure of the shocks formed and the evolution of the magnetic field in the shock using a 2-D magnetohydrodynamic simulation. As expected, the development of the magnetic field does not depend on the  $\beta$ value. Our results  show that instead of a single uniform shock forming behind the cosmic string we have multiple shocks forming at short time intervals behind the string. The presence of multiple shocks will definitely affect the observational signatures of cosmic string wakes as these signatures depend upon the temperature fluctuations generated by the shock. We also find that as the shock moves away, the residual magnetic field left behind reconnects and dissipates rapidly. The magnetic field around the string is thus very localized. We find that magnetic field reconnections take place in cosmic string wakes. This leads to the decrease of the magnetic field in the post shock region. } 

\end{abstract}
\begin{keyword}
cosmic string, wakes, magnetic reconnection 
\end{keyword}

\end{frontmatter}


\section{Introduction}

Magnetic fields form  an important part of our universe. Though magnetic fields of various magnitudes have been observed in various parts of the universe, the origin and evolution of these magnetic fields are still not clearly understood. The current observations suggest that the current magnetic field may be the result of the magnification of an initial seed magnetic field from the pregalactic era \cite{grasso,subramanian,hutschenreuter}. There are various mechanisms whereby the seed field could have been generated. One of the important methods for the generation of the seed magnetic field is the Biermann battery mechanism \cite{schoeffler}. This method will generate a magnetic field whenever the temperature gradient and the density gradient in an inhomogeneous plasma are not aligned to each other. Though this can happen under various circumstances in the early universe \cite{hanayama}, we are  interested in the magnetic fields that are generated in the wakes of cosmic strings \cite{sovan}.

Cosmic strings are one dimensional topological defects produced by the symmetry breaking phase transitions in the early universe. These strings can give rise to interesting cosmological consequences. They generate density fluctuations in the early universe. The density fluctuations behind a moving cosmic string generates what are called the wakes of a cosmic string.  The space-time geometry around a cosmic string is locally flat but globally conical. This leads to a planar overdensity behind the string. For a cosmic string moving with relativistic velocity, shocks will form in the cosmic string wake \cite{deruelle}. In relativistic fluid flows, both strong shocks as well as weak shocks can be generated \cite{hiscock}. The fluid flow around a cosmic string has been discussed in the literature \cite{stebbins,periva,zanchin}. Various signatures of these strong and weak shocks have also been discussed \cite{ponce} in the literature. Currently, there are various attempts to predict the signature of these cosmic string from the 21 cm observational data \cite{maibach}. 

 There are various ways in which these wakes can generate a seed magnetic field. Apart from the Biermann mechanism mentioned previously, magnetic fields can be generated close to the cosmic strings due to the Harrison mechanism \cite{harrison}. Two cosmic strings moving past one another will also generate a magnetic field \cite{vachaspati}. The seed magnetic field generated in the wake is usually quite small. It is assumed that the high Reynolds number of the early universe plasma will lead to vorticity in the cosmic string wake which will enhance the seed magnetic field. The main problem with these scenarios is that it is not clear whether the vortical motion generated by these wiggly strings can generate the large scale magnetic fields in the universe. This is mostly because the magnetic fields predicted by these models are quite weak ($B \approx 10^{-23} G$) \cite{avelinoshellard}. Larger fields may be produced if the strings are superconducting. These superconducting strings move slower than their non superconducting counter parts. It was estimated that they may generate magnetic fields as large as $10^{-19} $ G with a coherence scale of $1$ Mpc \cite{dimopoulos}. There have been some studies of shocks generated by superconducting cosmic strings with magnetized wakes \cite{chudnov}. It has been shown that bow shocks similar to the bow shocks in the sun's magnetosphere can be generated in the wakes of these superconducting strings. However, in this work we plan to study wakes generated by Abelian Higgs strings. These cosmic strings will  have magnetic fields generated in their wakes due to the Biermann battery mechanism.  

In recent times, numerical simulations have been developed to study the evolution of shock waves generated in various astrophysical scenarios. The microphysics of non relativistic as well as relativistic shocks are being studied using detailed magnetohydrodynamic simulations. Such studies have revealed interesting details about the evolution of magnetic fields in the different kinds of shocks. For details about recent studies in shocks please refer to ref. \cite{marcowith}. Though there are some numerical studies of wakes due to cosmic strings  previously \cite{stebbins,sornborger}, none of these studies have included the evolution of the magnetic field in the cosmic string wake.

In this work, we have studied the evolution of the magnetic field in the planar wake of a cosmic string. As the string moves, it leaves behind a wake in the plasma, we primarily study the wake left behind by the cosmic string. 
For a plasma having a magnetic field, the quantity $\beta$ gives the ratio of the thermal plasma pressure to the magnetic field pressure. The $\beta$ value in the early universe is generally very high even though the seed magnetic field may be smaller \cite{chirakkara}. We study the evolution of the wake of a cosmic string in a high $\beta$ plasma using the resistive MHD equations based on the OpenMHD code. We are interested to find out the parameters which determine the evolution of the magnetic field in the cosmic string wake. We find that several small shocks are generated and the magnetic field is high close to the cosmic string. While for lower values of $\beta$, the magnitude of the field depends on the $\beta$ value, this gets saturated at higher values of $\beta$. As the string moves through the plasma, it leaves behind magnetic field loops which reconnect. The magnetic field is therefore highest close to the shockfront and decreases in the post shock region. 

In section 2, we will review the formation and evolution of cosmic string wakes in the literature. In section 3, we will describe the details of the numerical set up that we have used to model the wakes of the cosmic strings in this work. In section 4, we will present the results that we have obtained from our simulations. In section 5, we will summarize and present the final conclusions.    

\section{Wakes due to cosmic strings}

Cosmic strings move with a high velocity in the early universe. The velocity of a cosmic string is close to the speed of light. Density fluctuations are generated behind a moving cosmic string. These are known as cosmic string wakes. As mentioned before, the space-time geometry around a cosmic string is locally flat but globally conical. If a string moves through the plasma it sweeps out a surface area just behind it due to the conical geometry of its space-time. A velocity perturbation is generated in the nearby matter due to the motion of the string. As the matter moves towards the region behind the cosmic string from both sides, the matter density will increase in that region and it will form a  two-dimensional structure. This two dimensional structure is a conical shape whose opening angle is given by the deficit angle of the cosmic string metric.

The metric around a cosmic string has a deficit angle given by $\delta \theta = 8 \pi G \tilde{\mu}$, where $\tilde{\mu}$ is the mass per unit length of the string. As the string moves forward, an observer behind the string would see matter streaming past it.  If the string moves with a velocity $v_s$ in a particular direction in a plane, the particles moving along that plane will get a velocity perturbation $\Delta v$ due to the deficit angle of the string.  The particles thus feel a velocity kick towards the center of the plane behind the string. The magnitude of the kick is given by $\delta v \sim \delta \theta v_s \gamma_s $ where $v_s$ is the velocity of the string and $\gamma_s$ is the relativistic factor. As more and more particles are kicked towards the string, an overdensity or wake is generated behind the string. A detailed description of wake formation is given in ref. \cite{silk}. Strings generated in the early universe generally move at relativistic velocities. Long strings moving at such high velocities may get chopped into smaller loops. So a long string moving at a time $t_{i}$ will generate a wake whose dimensions are given by $c_1 t_i \times t_i v_s \gamma_s \times \delta \theta  t_i v_s \gamma_s $. Here $c_1$ is a constant of order one \cite{brandenberger}. The opening angle of the wake depends upon the deficit angle of the cosmic string. 

The overdensity in the wake will lead to further accretion of matter, and the wake will grow in thickness. Generally, the plasma is charge neutral and hence the overdensity too is charge neutral. However as mentioned in the introduction it is possible that seed magnetic fields are generated in the wake of the string. In previous studies, the seed magnetic field generated is usually small compared to the equipartition field at the particular temperature. 
So for the magnetic fields to survive it is assumed that these fields get amplified in the cosmic string wakes. As such there has been no detailed study of the evolution of magnetic fields in the wakes of cosmic strings. We are therefore interested to see how a  magnetic field generated close to the cosmic string will evolve as the cosmic string moves through the plasma.

The simulation of turbulent plasma dynamics is non trivial and requires certain assumptions. As a first step, we simulate the shock structure in the cosmic string wake and do not include the dynamo effect in our current simulation. The production of shocks behind cosmic string wakes has been predicted before \cite{deruelle,stebbins}. We generate a shock wave behind the cosmic string and study the evolution of the magnetic field in the shock wave. The length scale of the wake will be of the horizon size and the width will be a fraction of the length. The width is determined by the deficit angle of the string and is proportional to  $\delta \theta  t_i v_s \gamma_s $. Though the deficit angle of the cosmic string is of the order of $10^{-5}$, for our simulations we take a much larger value. This is done so that the wakes are easier to study.

\section{Numerical method used to simulate the cosmic string wake}
\subsection{Boundary conditions and magnetohydrodynamic equations}
Since the space time around the cosmic string is locally flat and globally conical in nature, we model the inflow velocity as a  flow around a cone centered around the cosmic string. This means that the computational domain (spatial) would be making an angle $\theta$ with respect to the standard  $x-y$ plane. The domain will also be axis symmetric about the $y$ axis. We consider the cosmic string to move from left to right then the plasma particles streaming past it would be deflected by an angle $\theta$. The conical geometry of the string will result in the particles forming an overdensity behind it.     
This means that our initial velocity is given by, 
\begin{equation}
v_x = v_0 cos\theta ~~ ;  ~~
v_y = v_0 sin\theta  
\end{equation}
where $v_0$ is the initial velocity and $\theta$ is the opening angle of the wake. The initial velocity is actually a combination of the string velocity and the random velocity of the particles in the plasma. This means that the initial velocity cannot strictly be a  constant. However, since the cosmic string drags the plasma particles causing the wake, we are assuming that the string velocity dominates over the particle velocity. So we  take the initial velocity as a constant.  The wake structure with this boundary condition  is given in Fig. 1. The position of the cosmic string in fig 1 is at x = 550. We have shown it in this position so that it has a well developed wake structure around it.  In our initial conditions we have given the initial velocity of the particles and the magnetic field in the plasma around the moving cosmic string. There are no other initial conditions imposed on the differential equations to be solved in the code. We do not have any reflections at the boundary and the computational boundary is an artificial boundary through which the solution holds without any distortion. So even when the string passes through the boundary the wake structure remains behind and the continuity of the solution is maintained.

We have used the OpenMHD code to simulate the wake behind the cosmic string. The OpenMHD code is a finite volume code with Riemann solvers (also known as Godunov codes) developed by Zenitani et. al.\cite{zenitani}. We chose this code as it has the ability to describe discontinuities. This makes the code useful for studying shock-like structures. The time evolution in this code is handled by a Runge Kutta method and the simulations are carried out in the x-y two dimensional plane. We have plotted a lattice size of $x = 100$ to $x=900$ to study the structure of the wake. In the $y$  direction the plot is upto $ y = 150 $. on both sides of the y - axis.  The OpenMHD codes solves the following resistive MHD equations \cite{zenitani} (subject to the given boundary conditions),
\begin{align}
\frac{\partial {\rho }}{\partial t} + \vb*{ \nabla}. ( \rho \vb*{v} ) = 0 \\
\frac{\partial {\rho \vb*{v} }}{\partial t} + \vb*{ \nabla}. ( \rho \vb*{v} \vb*{v} + p_T \vb*{I}- \vb*{B} \vb*{B}) = 0 
\end{align}
\begin{align}
\frac{\partial {e}}{\partial t} + \vb*{ \nabla } .(e + p_T) \vb*{v}- (\vb*{v} .\vb*{B}) \vb*{B} + \eta \vb*{j} \times \vb*{B} = 0 \\
\frac{\partial {\vb*{B}}}{\partial t} + \vb*{ \nabla}. (\vb*{v} \vb*{B} - \vb*{B} \vb*{v})+ \vb*{\nabla} \times (\eta\vb*{j}) = 0. 
\end{align}
Here,the total pressure $p_T$ is given by $ p_T= p+ \frac{B^2}{2} $ and the total energy density by $e=\frac{p}{\Gamma-1}+\frac{\rho v^2}{2}+\frac{B^2}{2} $. A polytropic equation of state is considered with $\Gamma = \frac{5}{3}$. The OpenMHD is available in many versions for the simulation of different structures in the early universe. There are versions which involve a charge sheet, these are used to study magnetic reconnections. However, we have used only the basic structure of the code and modified the model part to suitably reflect the magnetic fields in the cosmic string wakes and the conical spacetime around the cosmic string.

\begin{figure}
	\includegraphics[width = \linewidth]{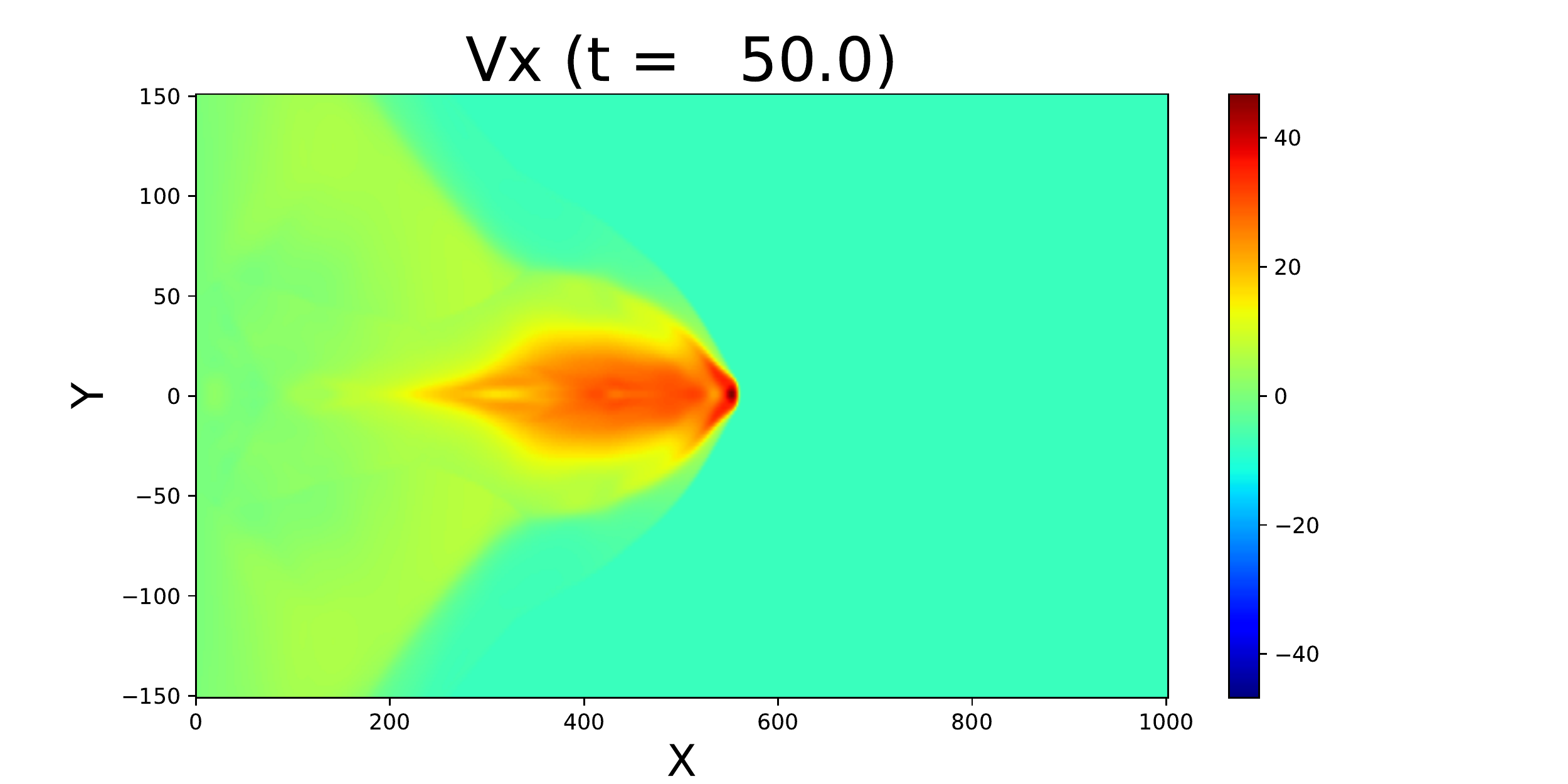}
	\caption{The wake structure due to a moving cosmic string }
	\label{fig:initialconfig}
\end{figure}

We are interested in the motion of the string in the early universe, so we consider the timescale to be close to the recombination time. At this time, the plasma has a small but finite resistivity given approximately by $1.6 \Omega $cm \cite{lattanzi}.  Hence we do not make any modifications to the resistive MHD equations. 
This essentially means that the magnetic field lines are no longer frozen in the plasma and the magnetic field may diffuse in the plasma.

It has generally been seen that the magnetic field generated in the wake of a cosmic string has a value less that the calculated value of the equipartition field at that temperature. In such cases, the fluid in the early universe has a pressure that is higher than the magnetic pressure density. Their ratio is given approximately by, $\frac{B^2}{8 \pi p} \sim 3 \times 10^{-25} Gauss^2$ \cite{barrow}. However this will lead to a very high $\beta$ value of the plasma. The $\beta$ value of the plasma is the ratio of the plasma pressure density to the magnetic pressure density. We have considered this to be of the order of $10^{5}$ or less. The plasma $\beta$ value is dimensionless. For convenience, the different quantities in the equations (pressure, density, resistivity and magnetic field) are also made dimensionless.
For MHD equations, this is done by  scaling the pressure and the density by a given factor. The dimensionless plasma mass density is therefore $\frac{\rho}{\rho_0}$ while the dimensionless pressure is given by $\frac{p}{\rho_0 v_a^2}$, Here, $v_a$ is the Alfven velocity. The Alfven velocity for our simulation is $v_A \sim 0.32$. Since we want the  plasma density at the initial time to be approximately $4 \times 10^{-3} eV^4$, the scaling factor considered is $4.8 \times 10^{-7}$.  The dimensionless pressure and density at the initial time is therefore of the order of  $10^{5}$. This would give us the approximate plasma density close to re recombination scale. 

The magnetic field is non zero at the start of the simulation. Since the nature of the magnetic field in the wakes is not known we consider a magnetic field that is oscillatory in nature and has the maximum amplitude close to the cosmic string. It is  given by $B = B_0 \exp(-A y) cos(\omega y) \hat{y} $. The form is motivated by the nature of the magnetic field generated in ref.\cite{sovan} and the symmetry of the problem. The magnetic field generated is perpendicular to the direction of motion of the cosmic string. It would have been better to do a three dimensional simulation but since we are only looking at a two dimensional simulation, we have just kept the magnetic field perpendicular to the motion of the cosmic string. As mentioned before, there is a reflection symmetry about $y = 0$ and the magnetic field is therefore maximum at $y = 0$.  The constants $A$ and $\omega$ are taken to be $0.2$, $0.8$ respectively. The magnetic field is also scaled to make it dimensionless. To obtain the actual field, we need to multiply by $10^{-9} eV^2$, this will give us the magnetic field in terms of nano Gauss ($nG$). Magnetic fields in this range has been predicted in the recombination era \cite{vachaspati2} . The resistivity of the plasma is taken to be of the order of $\frac{1}{1000}$. It is also a dimensionless quantity and is obtained by scaling the resistivity by a factor of $10^{-9} sec$.

The simplest solution to the resistive MHD equation is an uniform motion. As the string passes through the plasma, the equilibrium is disturbed. So we consider 
small  perturbations to the equilibrium state. The equilibrium magnetic field is given by  $B = B_0 \exp(-A y) cos(\omega y) \hat{y} $, to this a perturbation $B_y'$ is added. The amplitude of the perturbation $B_y'$ is taken to be two orders of magnitude less than the equilibrium amplitude. So if the magnitude of the equilibrium field is $B_0 = 1$, the amplitude of the perturbation is of the order of $0.01$. We have checked for small variations in the frequency of the equilibrium magnetic field in our simulation but we find that these variations do not change the nature of our results. Other details of the OpenMHD code can be obtained from \cite{zenitani} and references therein

\subsection{Time and length scales in the simulation }

We are looking at the cosmic string wakes in the recombination era. The horizon length is of the order of $200$ Mpc. The width of the wake will depend on the opening angle. Since the deficit angle for the strings are very small, the width of the wake will be far smaller than the length of the wake. It will be of the order of $10^{-3}$ Mpc for a cosmic string generated in the early universe. As mentioned before we have taken the opening angle to be larger in our simulations so that the wake structure can be studied in detail. So if we scale it back to the sizes more relevant to the recombination era, one unit of the grid cell will  correspond to $\approx 10^{-5}$ Mpc. It has been shown that magnetohydrodynamic shocks in plasma with high conductivity have a width which is several mean free paths of the undisturbed gas particles \cite{marshal}. At the recombination era, the charged particles are the electrons and the protons. Due to the larger mass of the proton it is considered to be nearly stationary as compared to the electrons. The electron's mean free path with respect to the magnetic field is less than $10^{10}$ cm \cite{tajima}. So the width of the shock is several times the mean free path of the electrons. Due to the smallness of the deficit angle the shock wave is more like a jet and the pressure density and temperature changes suddenly. The conductivity determines the nature of the magnetohydrodynamic shock, if the conductivity was lower the shock width would have been wider and more diffuse. 

Typical shocks from cosmic strings post recombination can be of the order of $5$ Mpc \cite{vollick}, however, in our case the thickness of the shock region is smaller. We also use this lengthscale to define the Alfven transit time in this system.  The characteristic length scale is $l \sim 10^{-3} $ Mpc. So the Alfven transit time given by $\frac{l}{v_A}$ is used to normalize the time in the simulation. A single timestep in the simulation will thus correspond to $3 \times 10^{-3}$ Mpc. We use a fixed timestep for the evolution of the equations. Though we are looking at the shock over a short period of time, we find that there are interesting features in the shock behind the cosmic string. The conical space time poses some challenges close to the boundary. For certain time steps, at a few points non-physical solutions are seen close to the boundary but overall there is no instability in the plasma. We have also varied the initial conditions and find that the results are independent of the numerical values of the initial conditions. Usually for complex geometrical conditions, the problems arise when the choice of the time step is not taken properly. We have checked the simulation for different timesteps and given the results for the cases where the solutions reflect the plasma conditions in the early universe. 

\section{Results}
\subsection{Shocks in cosmic string wakes}

\begin{figure}
	\includegraphics[width = \linewidth]{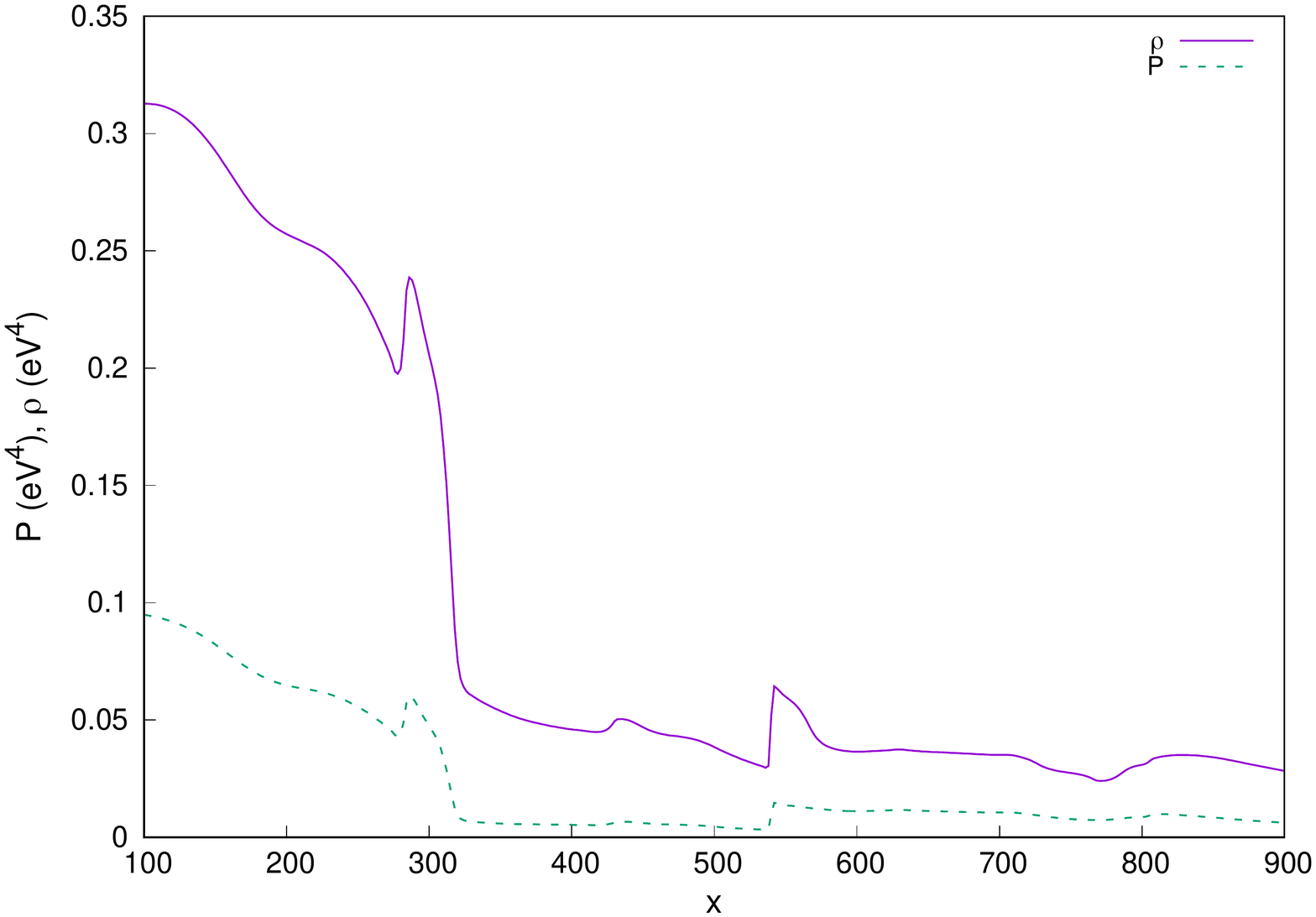}
	\caption{Plot of the pressure and density along the x-axis at $t = 140$ steps. The sharp drop in both the pressure and density  indicates the presence of a shock.}
	\label{fig:pressuredensity}
\end{figure}

As mentioned in the previous section, we are interested in the shock structure behind the cosmic strings. As the strings move through the background plasma, we plot the resulting pressure and total energy density at a particular time $t$ in figure \ref {fig:pressuredensity}. As is seen from fig. \ref {fig:pressuredensity}, there is a sharp rise in the density and the pressure simultaneously indicating the formation of a shock behind the string. This is the shock front and it moves against the flow of the plasma.  At the initial timesteps there is only one such sharp drop in the pressure and energy density but as the string moves through the plasma, multiple shocks can be identified at later times. The sudden change in the pressure and density is attributed to the jump conditions across the shock \cite{rankinehugoniot}. The figure is taken after $140$ timesteps. Due to the conical structure of the spacetime around the cosmic string, the flow close to the cosmic string is unsteady. However, as the system evolves, the system reaches a quasi - steady state. The figure \ref{fig:pressuredensity} reflects the quasi-steady state of the system.  

\begin{figure}
	\includegraphics[width = \linewidth]{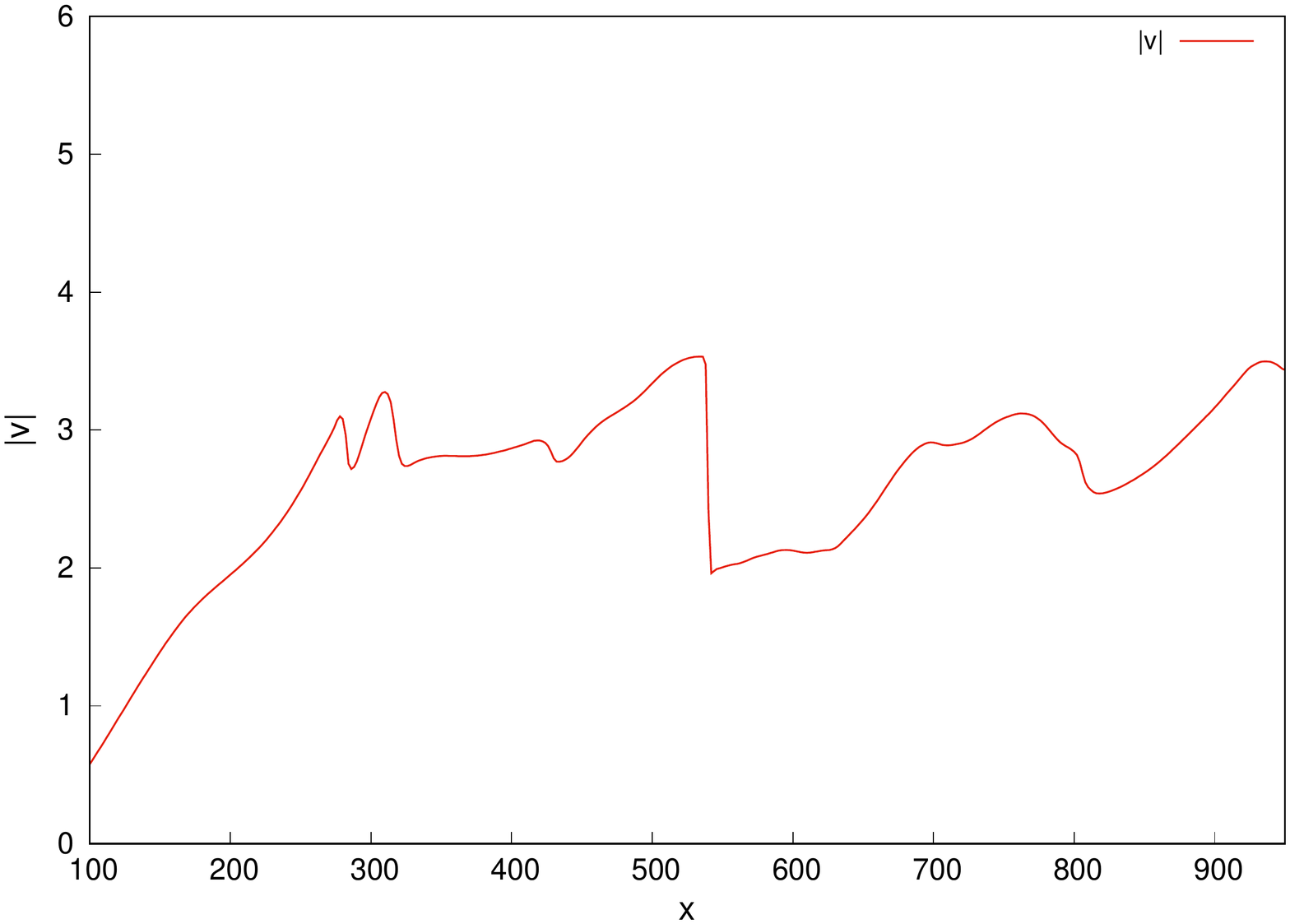}
	\caption{Plot of the velocity (magnitude only)  along the x-axis at $t = 140$ steps. We find a sharp change in the velocity at the same points as the pressure and density. The velocity is measured in terms of the Alfven velocity and is therefore dimensionless.}
	\label{fig:velocitypressure}
\end{figure}

Just as there is a sharp rise in the pressure and the energy density the total velocity also changes sharply close to the shock. This is shown in fig \ref{fig:velocitypressure}. The particle number density also changes across the shock front. It is typically of the order of $\frac{\delta \rho}{\rho} \sim 4 $. This will give rise to a temperature fluctuation. In ref.\cite{beresnyak}, it was shown that that shocks with $\frac{\rho_2}{\rho_1} \sim 4 $ can lead to detectable temperature fluctuations. However, the temperature fluctuations also depend upon the Mach number $M$ of the system. The system is supersonic and $M \sim 1 $ in these simulations. So the temperature fluctuations generated will be smaller than those predicted in ref.\cite{beresnyak}. For a wiggly cosmic string, the ratio of the post-shock temperature ($T_2$) to the preshock temperature ($T_1$) at recombination is given by, 
\begin{equation}
\frac{T_2}{T_1}~ =~ \frac{1}{2} ~ \Gamma ~(\Gamma - 1) ~(4 \pi G \tilde{\mu}  M )^2 
\end{equation}  
As the value of $ 4 \pi G \tilde{\mu} $ is small, this means that the temperature fluctuations from such strings will  be important for high Mach numbers. This can only happen for strong shocks. For $M \approx 1$, the temperature fluctuations will be smaller. However, the fact that there are multiple shocks following one another would mean that the lengthscales of the temperature fluctuations would also become important. Moreover unlike the case of steady shocks where the temperatures cool down as the shock passes by there will be consistent reheating of the plasma as  multiple shocks are generated as the wakes moves forward.

The plots clearly show that the shocks are non uniform and will therefore generate vorticity. Stream plots also show the generation of vorticity in the plasma. However the vorticity generated by a single stream in not enough to generate turbulence in the plasma. 

In magnetohydrodynamic shock waves there are multiple lengthscales involved. Apart from the lengthscale associated with the pressure and density changes, the other lengthscale would be given by the motion of the charged particles. The overdensity is mostly due to the heavier baryons (protons) whereas the charge is carried by the lighter electrons here. So for these magnetohydrodynamic shock waves the temperature fluctuations are not on the same lengthscale as the density fluctuations \cite{marshal}. Usually as the string moves forward, it leaves behind both density as well as temperature fluctuations. The lengthscale of the temperature fluctuations is usually given by the electron lengthscales as the electron temperature dominates the temperature fluctuations \cite{ponce}.
In a high $\beta$ and low Mach number plasma, the electron scale fluctuations and the ion scale fluctuations have different lengthscales. The electron length scale is given by, 
\begin{equation}
l_{elec} = \frac{1}{n_e \sigma_{T}}
\end{equation}
The ions (which in this case are assumed to be protons) have a lengthscale of, 
\begin{equation}
l_{pro} = \frac{1}{n_e \sigma_{T}} (\frac{m_p}{m_e})^2
\end{equation} 
 Here $\sigma_T$ is the Thomson cross section given by $\sigma_T = 6.65 \times 10^{-25} cm^2$. $n_e$ is the number density of electrons and $m_p$ and $m_e$  are the masses of the proton and electron respectively \cite{lesch}. The difference in lengthscales of the fluctuations have important consequence for the determination of the $21$ cm power spectrum. The power spectrum gives the correlations in the brightness spectrum as well as the cross- correlation between different fields of fluctuations \cite{furlaneto}.

 The observational consequences of these temperature fluctuations on the CMBR have been predicted previously \cite{brandenberger}. The density and temperature fluctuations are generally calculated based on the structure of a steady shock behind the cosmic string. For magnetized wakes, the presence of the different lengthscales indicates that the shock need not be a steady shock. Our numerical simulations show that multiple shocks can be generated in the magnetized string wake.  The formation of multiple shocks also indicate the possibility of shock collisions. This can also result in the generation of gravitational waves\cite{pen}.The possibility of having multiple shocks will mean that new observational signals may be obtained for cosmic strings with magnetized wakes.

\begin{figure}
	\includegraphics[width = \linewidth]{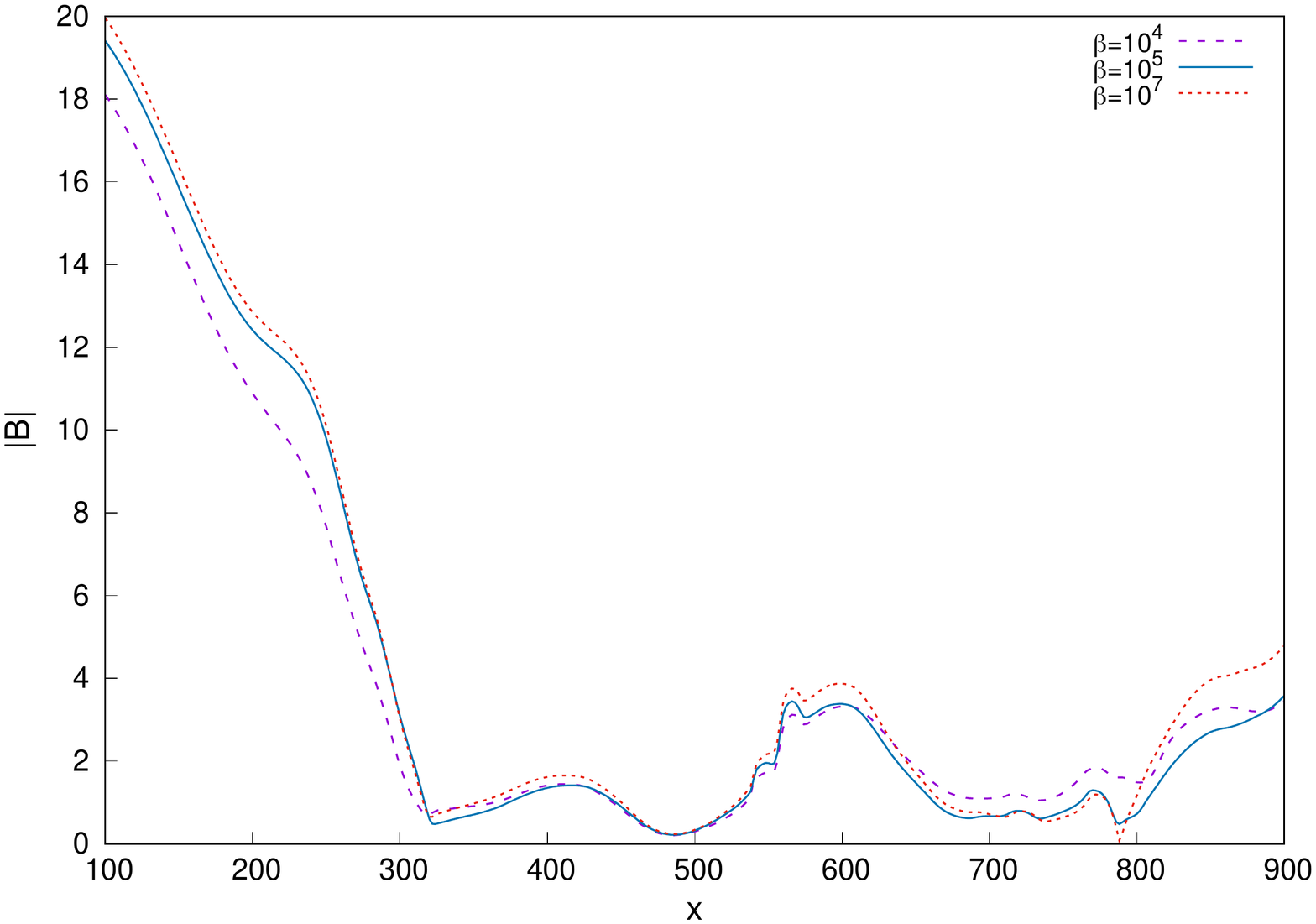}
	\caption{Plot of the magnetic field  $t = 140 $ steps for three different $\beta$ values.}
	\label{fig:betavalue}
\end{figure}

\begin{figure}
	\includegraphics[width = \linewidth]{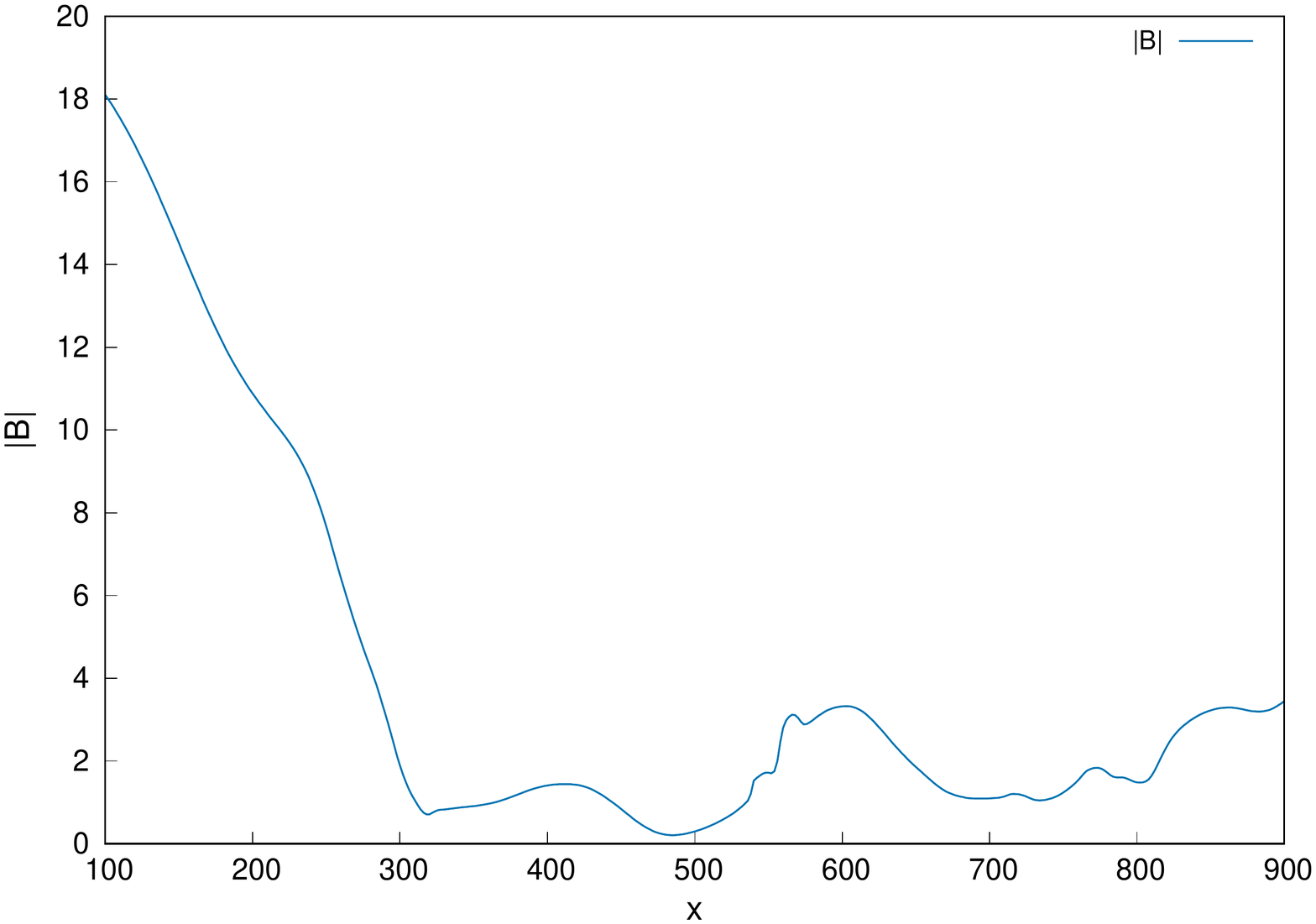}
	\caption{Plot of the magnitude of the magnetic field  at $t = 140$ steps. The shock is moving to the right. The plot gives the dimensionless value of the magnetic field.The dimension of the field can be obtained by multiplying with a factor of  $10^{-9} eV^2$ }
	\label{fig:magneticfieldspatial}
\end{figure}

We find that for high $\beta$ plasmas, the evolution of the magnetic field does not depend on the value of $\beta$ in the cosmic string wake. Since at the recombination era the $\beta$ value of the plasma is considerably high we take  $\beta \approx 10^4$ or higher. As is shown in  fig. \ref{fig:betavalue}, for all the different values of $\beta$, the plots are all similar.  
In fig. \ref{fig:magneticfieldspatial}, we plot the magnetic field at time $t= 140$. We find that the magnetic field is present over the whole range but its value is not uniform.   
We have taken several snapshots at several timesteps.
Since we have not included the dynamo effect in the simulations, we are not expecting the magnetic field to increase in magnitude. We see that without the dynamo effect, the magnetic field in the post shock region decreases. Hence it is important to have a dynamo mechanism to amplify the magnetic field in the cosmic string wakes. We hope to pursue this in a later work.

\subsection{Reconnection in the wakes of the cosmic string}

\begin{figure}
	\includegraphics[width = \linewidth]{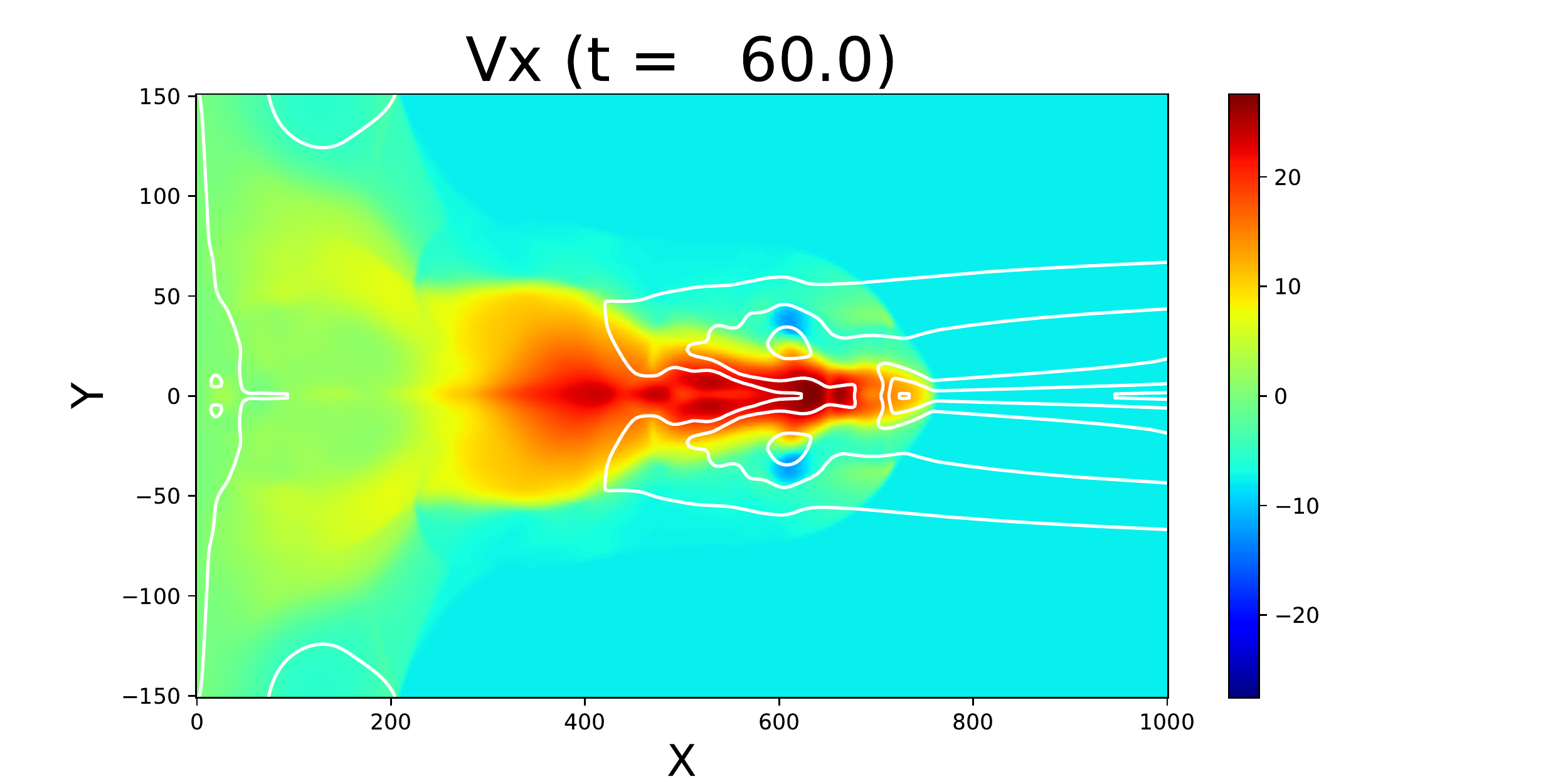}
	\caption{The wake structure due to a moving cosmic string with the magnetic field lines at $t = 60$}
	\label{fig:waket60}
\end{figure}

\begin{figure}
	\includegraphics[width = \linewidth]{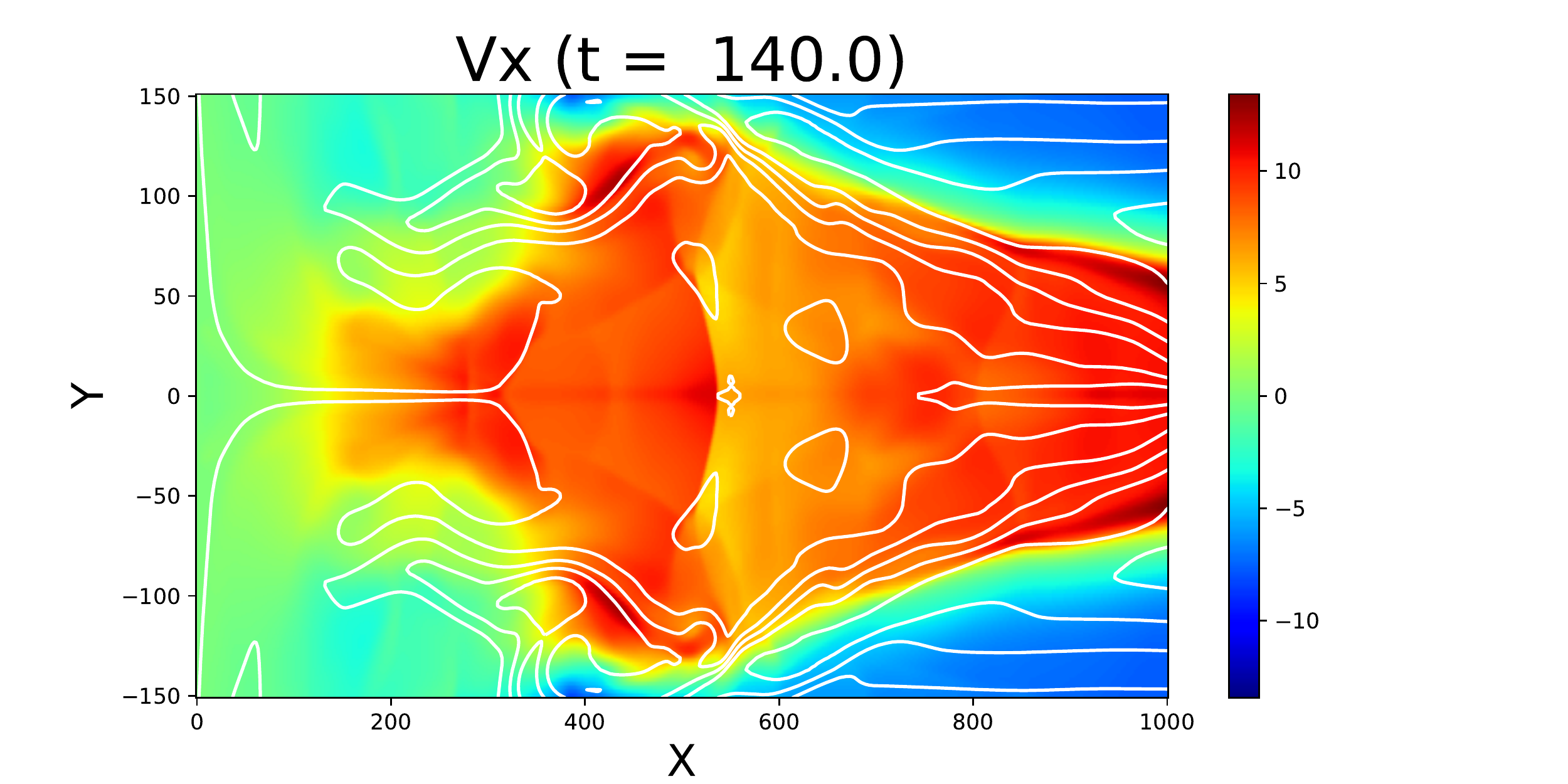}
	\caption{The wake structure due to a moving cosmic string with the magnetic field lines at $t = 140$}
	\label{fig:waket140}
\end{figure}

We have also plotted the magnetic fields in the wake of the cosmic string.  We would like to point out that there is a mirror symmetry in the code that has been used about the y-axis. 
So the lines of magnetic fields appear to be connected along the $y = 0$ line. 
Irrespective of this symmetry however, we see that the magnetic field lines do reconnect and form loops as the cosmic string moves away. This is seen as early as $t = 60 $ in fig \ref{fig:waket60}. The interesting part is that as the wake evolves we see many more magnetic field lines reconnecting and forming circular loops. So for $t = 140 $ in fig \ref{fig:waket140}  we see the formation of mutiple loops. 
Such magnetic reconnections occur due to small scale local electron dynamics. Anisotropic electron dynamics usually drive these reconnections. While the change in the velocity around a shock is given by the diffusion length scale of the particles in the plasma, the change in the magnetic field line is driven by the diffusion lengthscale as well as the conductivity of the plasma. Since we have a low resistivity plasma, the electron drift causes a charged current. This current causes the magnetic field lines to stretch. Stretched magnetic field lines can break and reconnect to form magnetic loops \cite{alt,biskamp}. As detailed local dynamics of magnetic fields in cosmic string wakes have not been studied before, such magnetic reconnections have not been discussed in the literature previously. Magnetic reconnections are associated with the formation of local charge configurations known as charge sheets. These can evolve and form well defined structures. It would be interesting to study the different conditions under which such charge sheets can be formed in cosmic string wakes. 

We also give the color map of the field amplitudes of the different components of the magnetic field in the figs 8 and 9. The actual magnitude in terms of dimension can be obtained by multiplying  the dimensionless values by $10^{-9} eV^2$.  The presence of magnetic reconnection can be understood from the fact that we have pressure changes in the cosmic string wakes coupled with the fact that the generated magnetic field is perpendicular to the flow direction and is also oscillatory in nature. Magnetic reconnections and charge sheets usually form when there is the possibility of the spatial variation of the different magnetic field components i.e the presence of a $\frac{\partial B_y}{\partial x}$ term in the evolution of the magnetic field equation. Thus this is not specific to the choice of the form of our magnetic field. Even if the magnetic field in the cosmic string wake is generated due to the vorticity of charged particles in the wakes of the cosmic string, the magnetic field will be spatially varying \cite{dimopoulos}. So there will be the possibility of magnetic reconnection occurring in cosmic string wakes irrespective of the method of generation of the magnetic field.   
 \begin{figure}
	\includegraphics[width = \linewidth]{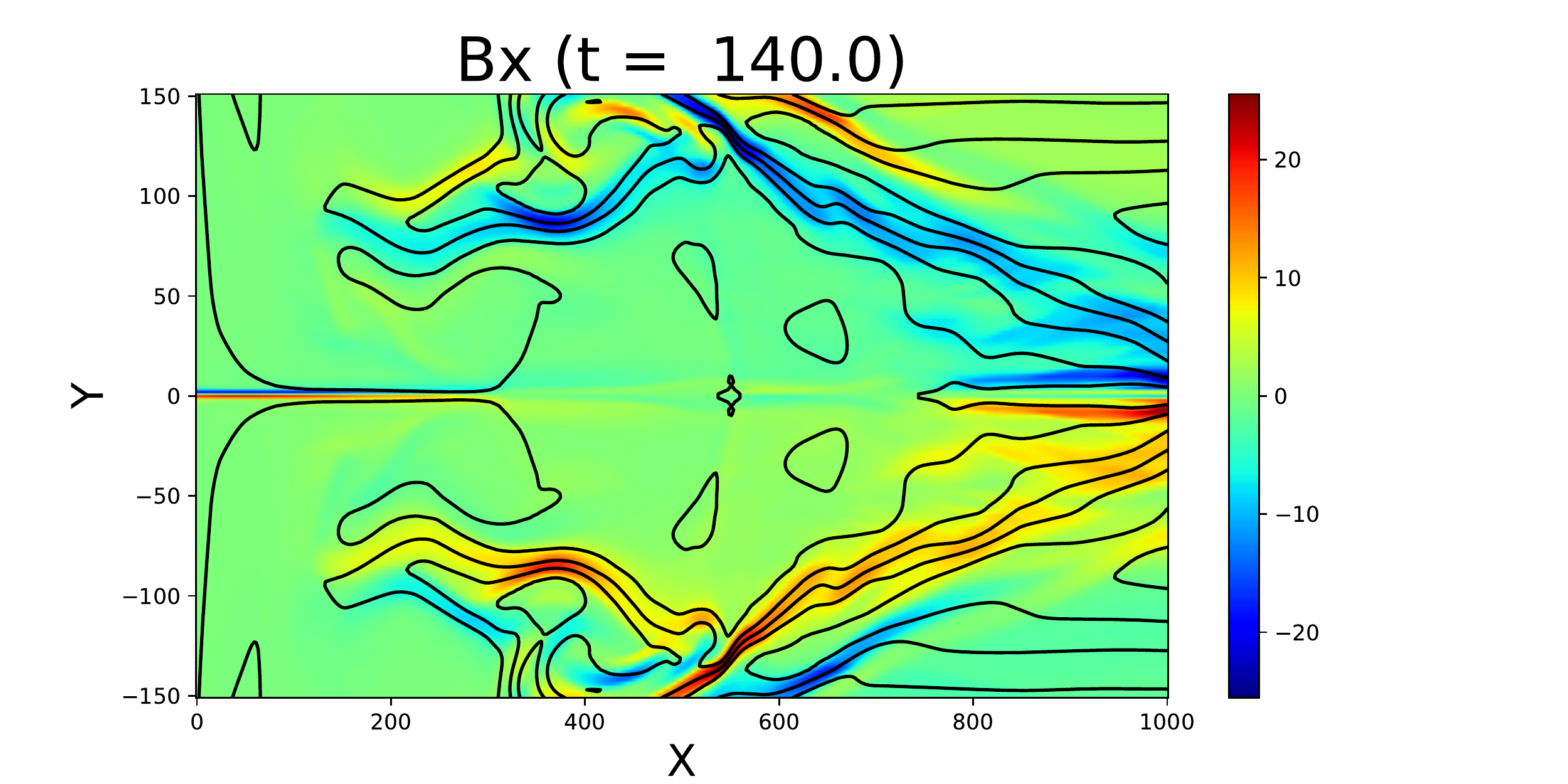}
	\caption{The surface plot showing the magnitude of the x-component  of the magnetic field at $t = 140$. The magnetic field in the figure is dimensionless. The dimension can be obtained by multiplying with $10^{-9} eV^2$. }
	\label{fig:magneticx}
\end{figure}
\begin{figure}
	\includegraphics[width = \linewidth]{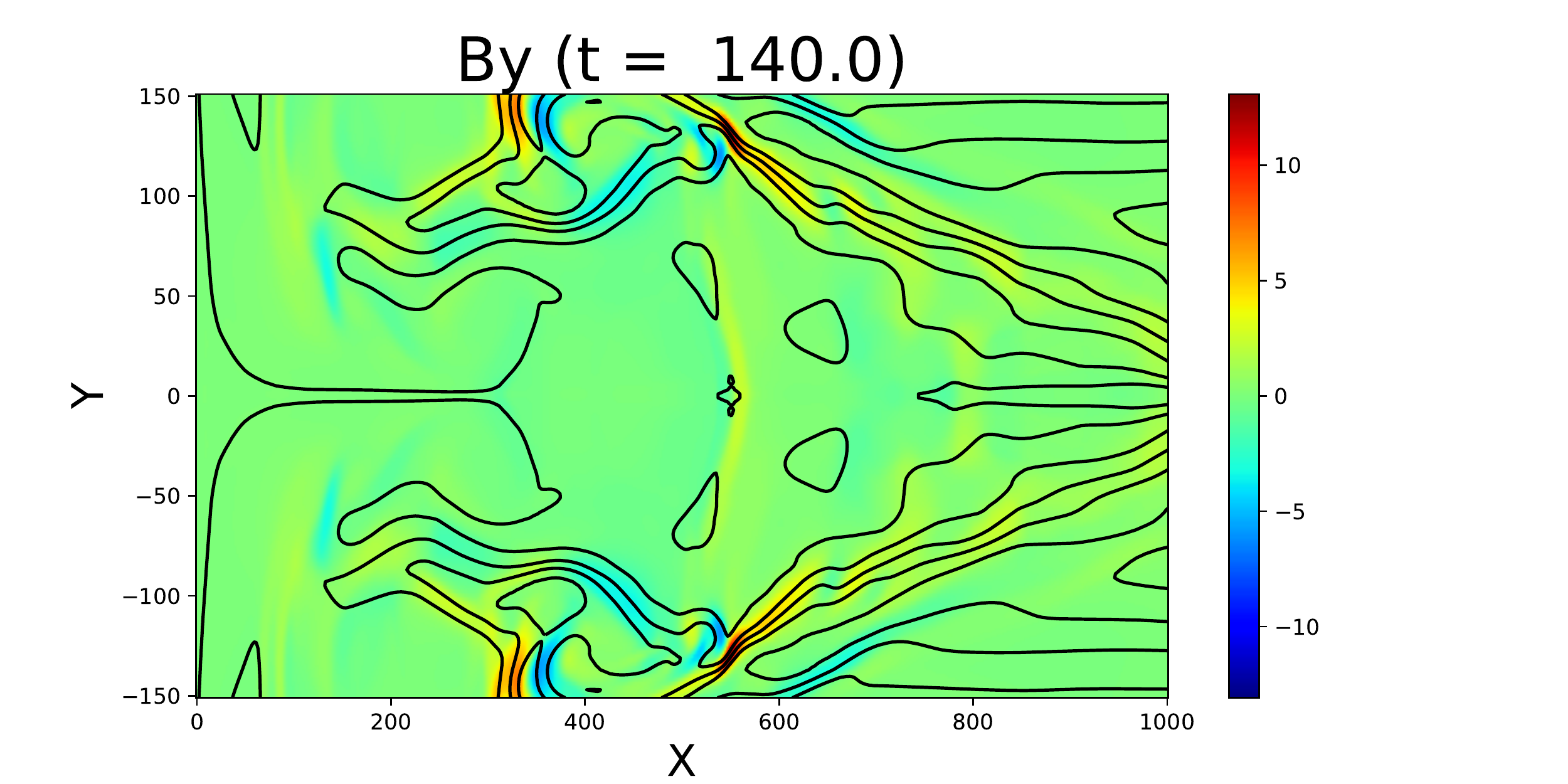}
	\caption{The surface plot showing the magnitude of the y-component of the magnetic field at $t = 140$. The magnetic field in the figure is dimensionless. The dimension can be obtained by multiplying with $10^{-9} eV^2$. }
	\label{fig:magneticy}
\end{figure}

\section{Summary and Conclusions}

In summary, we have done a detailed numerical study of magnetized shocks in cosmic string wakes. We find that it is possible to generate multiple shocks as the cosmic string moves through the plasma. Several observational signals have been predicted for shocks generated by cosmic string wakes, they are mostly for steady shocks. In this work, we have shown that the shocks generated in magnetized wakes need not be steady shocks. Depending on the nature of the magnetic field structure in the wake of the cosmic string, multiple shocks are possible. This has several consequences for the 
observational signatures predicted for the cosmic strings. The temperature fluctuations and the density fluctuations are not on the same lengthscale as they would have been for steady shocks.
The density power spectrum and the density correlations will therefore be different from the temperature power spectrum and the temperature correlations. This means that the observational signals of magnetized cosmic string wakes will be different from the signals expected from un-magnetized wakes.  
It is also possible that these shocks collide with one another and generate gravitational waves (GW). This is a promising aspect as recent work indicates that there are interesting prospects of detecting GW from turbulent sources at higher redshifts using space based interferometers and pulsar timing arrays \cite{tina}.

We have taken the opening angle of the wake to be very large. However, in a realistic environment the opening angle will be of the order of $10^{-5}$ radians. So the wake will be squeezed and will have a length that is much larger than it's width.At the time of recombination a realistic wake will thus have a length of the order of $500$ Mpc and a width of the order of only $10^{-2}$ Mpc.  This means that evolving magnetic field will correspond to a thin line like structure. The magnetic field will then be squeezed in the wake.  We have also found that there is a possibility of magnetic reconnections in cosmic string wakes. This may become enhanced in a realistic environment.  The magnetic field surrounding a realistic cosmic string may not be detectable but the magnetic energy generated through reconnections in the wake region may lead to the acceleration of charged particles which may be detectable. These particles may contribute to the cosmic ray spectrum. It has been postulated before \cite{montemerle}  that one of the sources of cosmological cosmic rays could be a burst of energy at high redshift which leads to the acceleration of charged particles like the protons and the $\alpha$ particles.  We plan to do a more detailed study of magnetic field reconnection in a later work where we would like to study the acceleration of charged particles and check whether they can be one of the unknown sources of radiation in the early universe. The reconnection of magnetic fields also affects the primordial Li (lithium) and D (deuterium) abundances \cite{lu}. So the magnetic reconnection in cosmic string wakes can also be used to address the small discrepancies in the Li abundance.  Apart from this, the injection of hadrons to the post recombination plasma will also affect the Big Bang Nucleosynthesis (BBN) calculations \cite{reno}. It should therefore be possible to use the BBN calculations to put constraints on the magnetic fields in the cosmic string wakes. 

We also find that the magnetic field in the shocks is not amplified unless there is an explicit dynamo mechanism that causes the amplification of the field. We hope to do more detailed numerical simulations by including the dynamo mechanism in a later work where we can study the magnetic field amplification in cosmic string wakes.

\begin{center}
 Acknowledgments
\end{center}  
 For computational infrastructure, we acknowledge the Center for Modelling, Simulation and Design (CMSD) at the University of Hyderabad, where part of the simulations was carried out. S.N acknowledges financial support from CSIR fellowship No. 09/414(2001)/2019-EMR-I  given by the Human Resource Development Group, Government of India. The authors would like to thank Abhisek Saha for help with the numerical simulations and Dilip Kumar for discussions.

\end{document}